\documentclass[a4paper,11pt]{article}
\usepackage{pos}
\usepackage{macros}
\usepackage{booktabs}
\usepackage[utf8]{inputenc}	
\usepackage[T1]{fontenc}		
\usepackage[english]{babel}		
\usepackage[rflt]{floatflt}

\usepackage[tableposition=top]{caption}
\usepackage{marginnote}
\usepackage{defs}
\usepackage{xcolor}
\usepackage[firstpageonly=true,angle=0,hpos=18.2cm,vpos=5.8cm,text=MS-TP-24-03,vanchor=t,hanchor=l,fontsize=12pt, color=black]{draftwatermark}

\title{Towards charm physics with stabilised Wilson fermions}

\author[a]{P. Fritzsch}
\author[b]{J. Heitger}
\author[c]{F. Joswig}
\author*[b]{J. T. Kuhlmann}

\affiliation[a]{School of Mathematics, Trinity College, College Green, Dublin, Ireland}
\affiliation[b]{Institut für Theoretische Physik, Universität Münster,\\
	Wilhelm-Klemm-Straße 9, 48149 Münster, Germany}
\affiliation[c]{Higgs Centre for Theoretical Physics, School of Physics and Astronomy,\\
    The University of Edinburgh, Edinburgh EH9 3FD, UK}

\emailAdd{fritzscp@tcd.ie}
\emailAdd{heitger@uni-muenster.de}
\emailAdd{fabian.joswig@ed.ac.uk}
\emailAdd{j.kuhlmann@uni-muenster.de}

\abstract{We report on a first study towards the use of stabilised Wilson fermions in heavy flavour physics.
	In particular, we are interested in fixing the charm quark mass via various physical observables and to inspect cut-off effects arising from different choices. This is done on large-volume OpenLat ensembles with periodic boundary conditions.
    Two different ways of fixing the charm quark mass are explored, namely using the mass of the $D$- and $\etac$-meson as physical inputs.
    We furthermore give an update on our determination of the non-singlet axial current improvement coefficient $\ca$.
}

\FullConference{%
  
}
\newcommand{\mqq}{m_{\rm q}}

\begin{document}
\maketitle
\section{Introduction}
With this project, we aim at exploring the capabilities of lattice QCD with stabilised Wilson fermions (SWF) for applications in heavy-quark physics, motivated by the possibility of stabilisation leading to smaller cut-off effects. Our goal is to calculate hadronic matrix elements involving heavy quarks that contribute, for instance, to the phenomenology of leptonic and semi-leptonic decays. 

As usual, the first step is to fix the charm quark mass by some physical input, which  can be done in multiple ways. At the moment, we are mainly focusing on employing the physical $D$- and $\etac$-meson masses, and investigating the resulting cut-off effects.

\section{Setup and tools}
Stabilised Wilson fermions, as proposed in \cite{Francis2019}, evolve the traditional Wilson fermion formulation to being less prone to numerical instabilities encountered due to near-zero modes of the Wilson--Dirac operator. Stabilising measures include, but are not limited to, the usage of the stochastic Molecular Dynamics (SMD) algorithm in the production of the gauge fields as well as an exponentiated form of the even-odd preconditioned $\rmO(a)$ improved Wilson--Dirac operator
\begin{equation}
    D_{\rm ee}+D_{\rm oo} = (m_0+4)+\csw\frac{i}{4}\sigma_{\mu \nu}F_{\mu \nu}\quad\rightarrow\quad (m_0+4)\exp\left(\frac{\csw}{(m_0+4)}\cdot\frac{i}{4}\sigma_{\mu \nu}F_{\mu \nu}\right)\,.
\end{equation}
Since there are already first indications for this formulation of Wilson fermions to have smaller cut-off effects~\cite{Fritzsch2022, Francis2019}, there is reason to believe that they are also well-suited for heavy-quark physics.

The data base of our study consists of OpenLat\footnote{\url{https://openlat1.gitlab.io}} ensembles which employ stabilised Wilson fermions. Further investigations within the SWF setup can, e.g., be found in refs.~\cite{Basta2022, Cuteri2022, Ce2023}. An update on the progress of gauge field ensemble production has been given at this conference~\cite{Cuteri2023}. The ensembles used here are all at the $N_{\rm f} = 3$ flavour-symmetric point and have lattice spacings in the range of $a\approx 0.120\,\dots\,0.064 \,{\rm fm}$ with periodic boundary conditions. While an even finer lattice spacing is available, we currently exclude this from the analysis.
We also note that the critical mass point $m_0=m_{\rm crit}$ is not yet known for these ensembles.

\begin{figure}
    \centering
    \includegraphics[width=.75\linewidth]{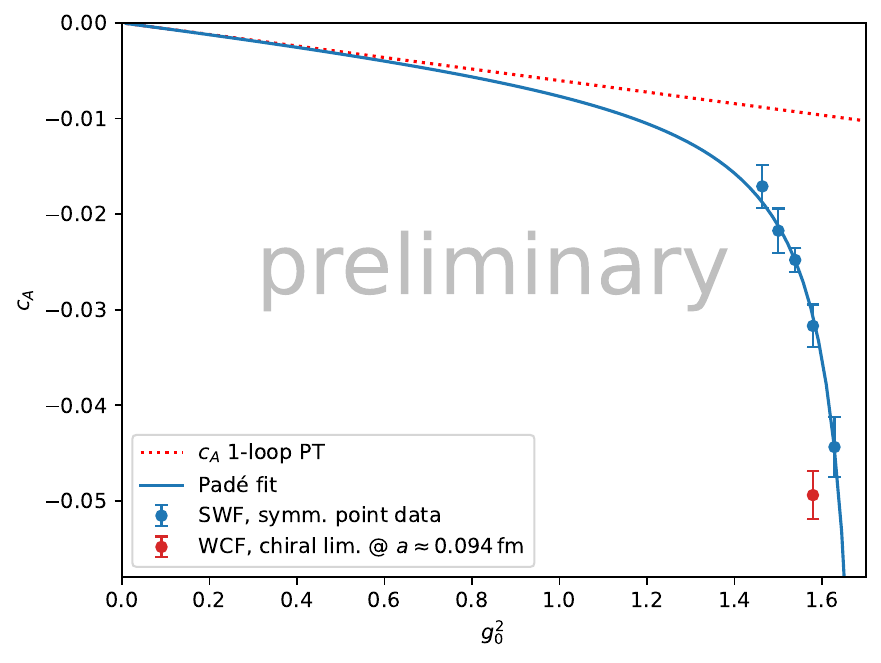}
    \caption{Preliminary interpolation formula~\eqref{eq:cA} for the improvement coefficient $\ca$ in the SWF framework. For comparison, the red point displays a result for the traditional non-perturbatively improved Wilson setup (WCF)~\cite{Bulava2015} at a matched lattice spacing of $a\approx 0.094\,\fm$.}
    \label{fig:ca_sym_point_plot}
\end{figure}

First estimates on the coefficient $\ca$, required to improve the axial-vector current
\begin{equation}
    A^a_{\mu,\rm I}(x) = A^a_\mu(x)+\ca a \partial_\mu P^a(x)
\end{equation}
 at $\rmO(a)$, have previously been reported in~\cite{Fritzsch2022}. It is used in our subsequent calculation of PCAC masses and matrix elements. Adopting the techniques introduced in~\cite{Morte2005, Bulava2015} at the flavour-symmetric point, we arrive at the preliminary empiric formula
\begin{equation}\label{eq:cA}
    \ca(g_0^2) = -0.006033\,g_0^2\,\left(\frac{1-0.4971\,g_0^2}{1-0.5823\,g_0^2}\right)
\end{equation}
modelling our numerical results as shown in fig.~\ref{fig:ca_sym_point_plot}.

Software wise, we are using \texttt{Hadrons}~\cite{Hadrons} built on the \texttt{GRID}~\cite{Boyle2015} back-end for the calculations presented here. The data and error analysis is done using the package \texttt{pyerrors} introduced in~\cite{Joswig2022}, which is based upon the Gamma-method~\cite{Madras1988, Wolff2003}.


\section{Fixing the charm quark mass}

To fix the mass of the charm quark, different procedures may be facilitated. 
In general, one takes a measurable quantity $X$, sensitive to the heavy quark's mass, and a (overall) scale-setting quantity $\bar{r}$, which (when appropriately combined) can be extrapolated to the continuum. Our intention is to probe different quantities to set the mass scale of the charm quark at finite lattice spacing. Without loss of generality, we take quantities $X_j\in \{ m_D, m_{D_\mathrm{s}}, m_{\eta_\mathrm{c}}, \ldots \} $ of mass-dimension $[X]_m=1$ and the Wilson-flow scale $\bar{r}=\sqrt{t_0}$ of mass-dimension $[\bar{r}]_m=-1$ to compute $\phi_j \equiv \bar{r}X_j$ on a given gauge ensemble, and demand that
\begin{align}\label{eq:mass-scale}
\phi_j(\Delta_h)|_{\Delta_h=\Delta_\mathrm{c}} \equiv [\bar{r}X_j]_\mathrm{phys} 
\end{align}
for some charm (valence) mass parameter $\Delta_\mathrm{c}$. In practise, $\Delta_h$ corresponds to a bare parameter of the theory, like the hopping parameter $\kappa_h$ or $m_{0,h}=1/(2\kappa_h)-4$, or any other parameterisation in one-to-one correspondence to those. One can equally well choose
any power of eq.~\eqref{eq:mass-scale} to fix this bare parameter. Here we choose $\bar{r}=\sqrt{t_0}$ because it is the quantity used to set the lattice scale on OpenLat ensembles as well as to fix the line of constant physics (LCP) for the light quarks.%
\footnote{Actually, $\sqrt{8t_0}$ has been used in the light quark sector. For the ease of presentation, however, we avoid the additional factor $\sqrt{8}$ such that in continuum extrapolations a common scale factor $\bar{r}/a$ enters on both axes.}
By imposing the condition~\eqref{eq:mass-scale} and evaluating it numerically, one obtains $\Delta_{\rm c}$.

For this purpose and the time being, we focus our attention on the ground-state mass of pseudo-scalar charmed mesons.
These are the $D$-, $D_\mathrm{s}$- and $\eta_\mathrm{c}$-meson masses. For the latter, one expects the best and most stable signal-to-noise ratio, as the charmonium channel is made of mass-degenerate quarks. We neglect the related disconnected contributions that are typically suppressed for heavy quarks in the loop. As physical input for the heavy-light pseudo-scalar mesons, $D$ and $D_\mathrm{s}$, we advocate their $SU(3)$ flavour-averaged mass including QED corrections,
\begin{align}\label{eq:DDs-avg}
    m_{D,{\rm avg}} = \frac{1}{3}(2m_{D} +m_{D_\mathrm{s}}) = 1899.40(37)\,{\rm MeV} \;,
\end{align}
based on masses listed in the PDG~\cite{PDG2022}, namely $m_{D}=1866.1(2)\,\MeV$ and $m_{D_s}=1966.0(4)\,\MeV$. This flavour average is motivated by the light quark LCP which keeps the trace of the $\Nf=2+1$ quark mass matrix approximately constant. Hence, one also expects a significant suppression of its pion mass-dependence. For the $\etac$-meson, on the other hand, the dependence on the sea-quark mass is naturally subleading and not taken into account explicitly. Accordingly, we take its face value directly from the PDG~\cite{PDG2022}, $m_{\etac} = 2983.90(40)\,\MeV$.

For different inputs $X_j\in \{m_{D, {\rm avg}}, m_{\etac}\}$, we now evaluate eq.~\eqref{eq:mass-scale} on four symmetric-point lattices with couplings $\beta=6/g_0^2\in \{4.0,3.9,3.8,3.685\}$, yielding $\Delta_\mathrm{c}=\Delta_\mathrm{c}(\beta,X_j)$.
\begin{figure}[t!]
    \begin{tabular}{cc}
         \includegraphics[width = 0.45\linewidth]{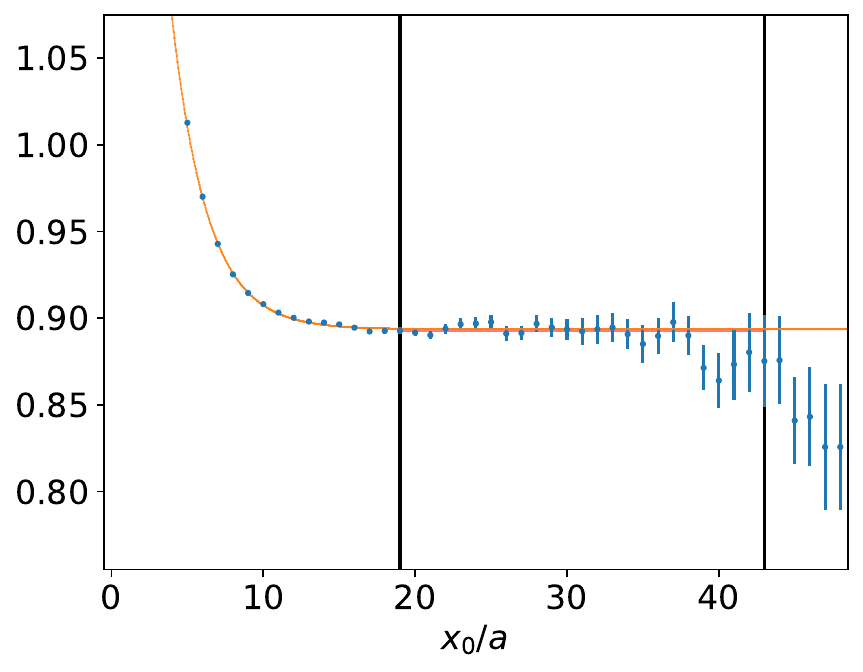}&
         \includegraphics[width=0.45\linewidth]{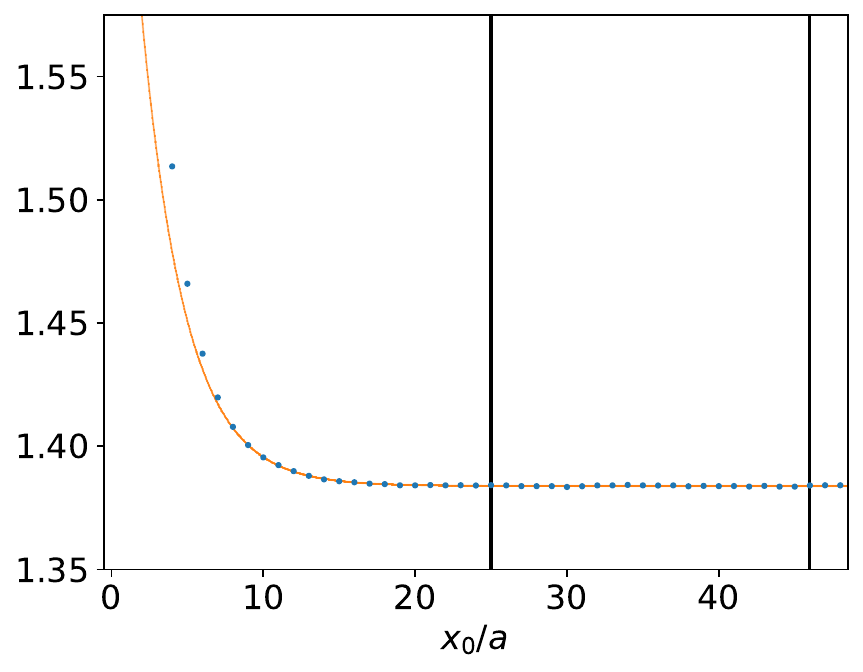}
    \end{tabular}
    \caption{Representative plateau in the mass of the $D_\mathrm{avg}$-meson (left) and the $\etac$-meson (right), shown here for an ensemble with lattice spacing $0.094\,{\rm fm}$, in vicinity of the estimated charm quark mass parameter $\Delta_{\rm c}$.}
    \label{fig:plat_mesons}
\end{figure}
\begin{figure}[t!]
    \centering
    \begin{tabular}{cc}
    \includegraphics[width=0.45\linewidth]{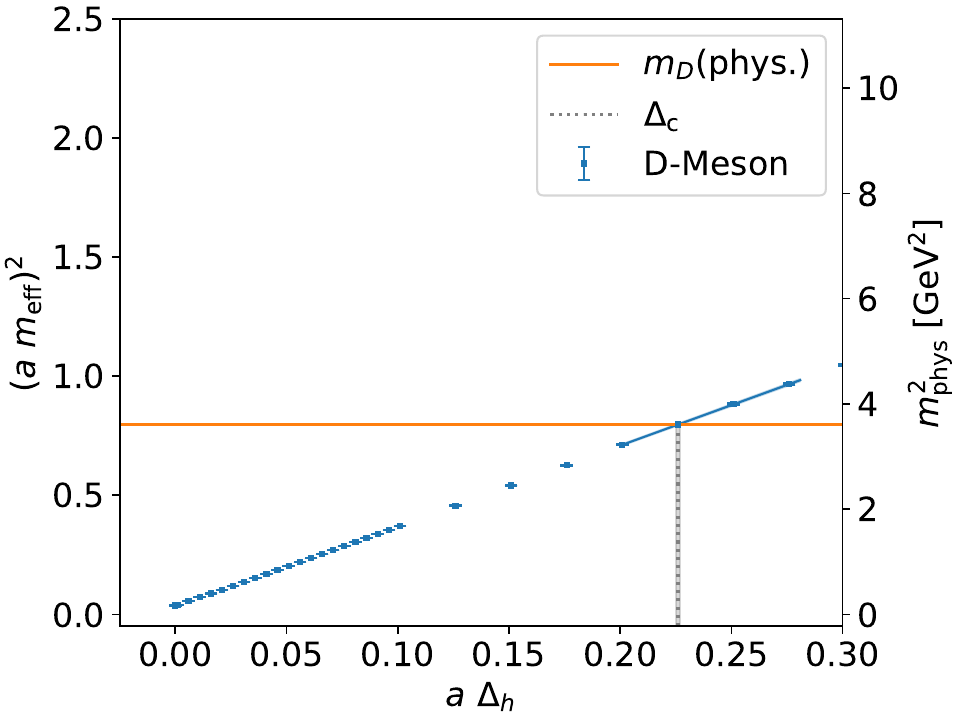}
    &
    \includegraphics[width=0.45\linewidth]{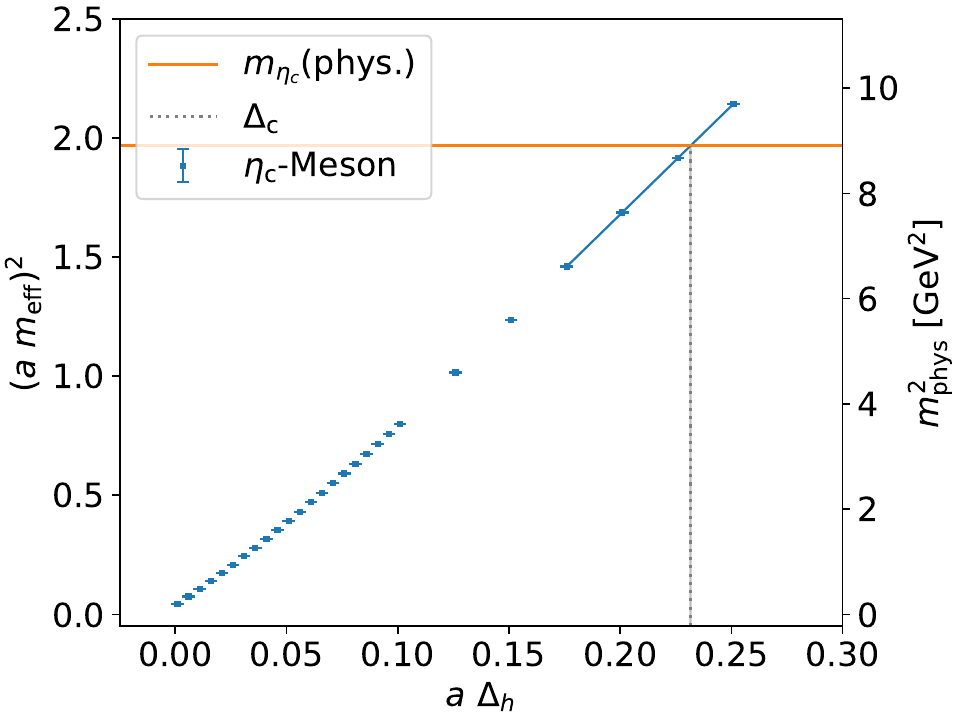}
    \end{tabular}
    \caption{Finding the charm quark mass parameter $\Delta_{\rm c}$ via matching the effective masses of the $D_{\rm avg}$-meson (left) and $\etac$-meson (right) to their physical target values.}
    \label{fig:find_charm}
\end{figure}
In order to find the point where the measured and physical meson mass are equal, a scan over a range of heavy quark masses is done and the corresponding effective masses are computed. Interpolating the plateau-averaged effective masses as a function of our (heavy valence quark) mass parameter
\begin{equation}
    \Delta_h = \frac{1}{2}\left(m_{0,{h}}-m_{0,{\rm sym}}\right) = \frac{1}{4}\left(\frac{1}{\kappa_{h}}-\frac{1}{\kappa_{\rm sym}}\right)
\end{equation}
to the target physical input mass fixes $\Delta_{\rm c}$ according to eq.~\eqref{eq:mass-scale}. In this way, $\Delta_h$ parametrises the difference to the unitary point ($m_{0,h}=m_{0,{\rm sym}}$), and the knowledge of $m_\mathrm{crit}$ is not required.

To identify these mass plateaux, the effective masses are fitted with a two-state fit of the form
\begin{equation}
    m_{\rm eff}(x_0) = c_0+c_1e^{-\delta_1x_0} 
\end{equation}
in which $c_0$ is the ground-state mass that we are interested in, while $\delta_1$ denotes the mass gap between the ground and the first excited state.
The starting point of a plateau is then defined as the time slice, where the contribution from the first excited state is 4 times smaller than the statistical error of the data, $\sigma_{\rm stat}(m_{\rm eff}(x_0))/4>c_1e^{-\delta_1x_0}$.
Since effective $D$-meson masses do not exhibit a stable plateau any more close $x_0\simeq T/2$, cf.~fig.~\ref{fig:plat_mesons} (left), our plateau average terminates (quite arbitrarily) at the time slice where $\sigma_{\rm stat}(m_{\rm eff}(x_0))>0.03\,m_{\rm eff}(x_0)$. Such a criterion is usually not needed for $\etac$-mesons, as they do not suffer from a similar signal-to-noise problem for large $x_0$ (see fig.~\ref{fig:plat_mesons} on the right).
Having determined various effective mass plateaux for different bare mass parameters in the neighbourhood of the physical value, we can locally linearise the functional behaviour to evaluate eq.~\eqref{eq:mass-scale} numerically; this is illustrated in fig.~\ref{fig:find_charm}.  

Besides the method to fix a desired physical quark mass discussed before, which is typically applied in large volumes (i.e., in the hadronic regime where finite-volume effects are exponentially suppressed), we also intend to perform a matching procedure that, so far, has only been used in small volumes. This amounts to first solve the renormalisation and improvement problem for the quark mass, which is quite involved for non-perturbatively improved Wilson fermions. For instance, a general bare subtracted quark mass, $m_{{\rm q},i}=m_{0,i}-m_{\rm crit}$, is renormalised and $\rmO(a)$ improved as
\begin{align}
    m_{i,{\rm R}}&= Z_{\rm m}\left\lbrace\left[m_{{\rm q},i}+(r_{\rm m}-1)\frac{\Tr[M_{\rm q}]}{\nf}\right]+aB_{i}\right\rbrace+\rmO(a^2)\\
    \hspace{0.3cm}
    B_i=\bm m_{{\rm q},i}^2&+\bar{b}_{\rm m} m_{{\rm q},i}\Tr[M_{\rm q}]+(r_{\rm m}d_{\rm m}-\bm)\frac{\Tr[M_{\rm q}^2]}{\nf}+(r_{\rm m}\bar{d}_{\rm m}-\bar{b}_{\rm m})\frac{\Tr[M_{\rm q}]^2}{\nf}\;, \nonumber
\end{align}
where $\Tr[M_{\rm q}]$ is the trace of the simulated (light) quark masses, here for $\Nf=3$. We note that for non-vanishing sea quark mass, $\Tr[M_{\rm q}]\ne 0$, there is a shift in the physical quark mass in the continuum, i.e., at $\rmO(a^0)$, and additional improvement coefficients appear in the $B_i$ counter-term. 

While established procedures relying on Ward identities exist to non-perturbatively determine $Z_{\rm m}$ and $b_{\rm m}$ that survive in the chiral limit, as well as $r_{\rm m}$, see refs.~\cite{Guagnelli2001, Fritzsch2010, Divitiis2019a,Heitger:2021bmg}, similar approaches have not yet been explored to calculate $\bar{b}_{\rm m}$, ${d}_{\rm m}$ and $\bar{d}_{\rm m}$. The $d$-coefficients require a variation of the singlet (sea quark) mass combination but multiply a term quadratic in the light quark mass and may be neglected in a first attempt. The general strategy to determine these coefficients is via appropriate improvement conditions that usually mimic some derivative (or linear combinations thereof) which is sensitive to a particular coefficient. Note that the improvement pattern simplifies when quark mass differences are considered, but lacking knowledge of the remaining coefficient $\bar{b}_{\rm m}$ (as well as of
the critical mass point $m_{\rm crit}$ on a few ensembles) for now, exploiting this promising option to fix the charm quark mass scale is left for the future.

\section{Preliminary scaling tests}
\begin{figure}
    \begin{tabular}{cc}
         \includegraphics[width = .45\linewidth]{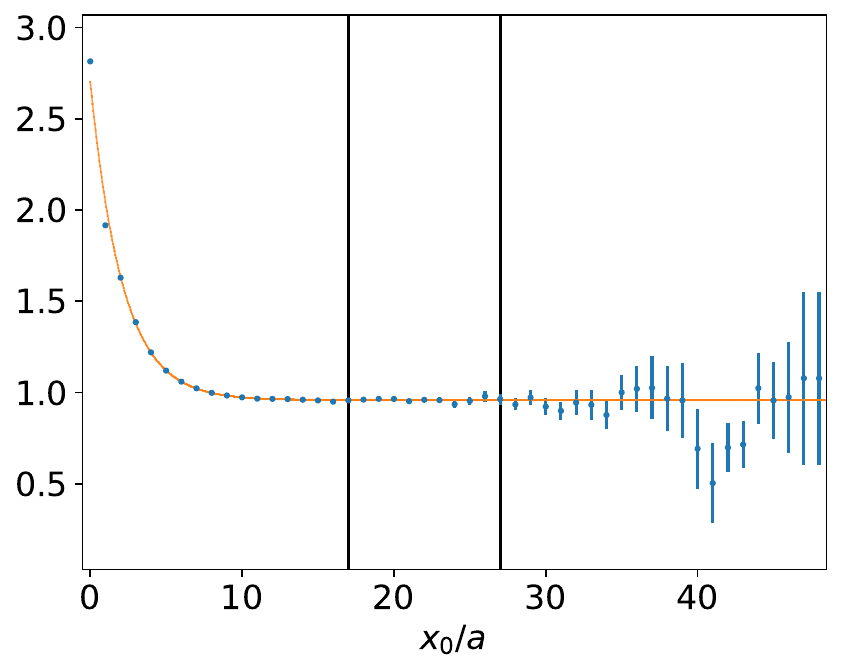}&
         \includegraphics[width = .45\linewidth]{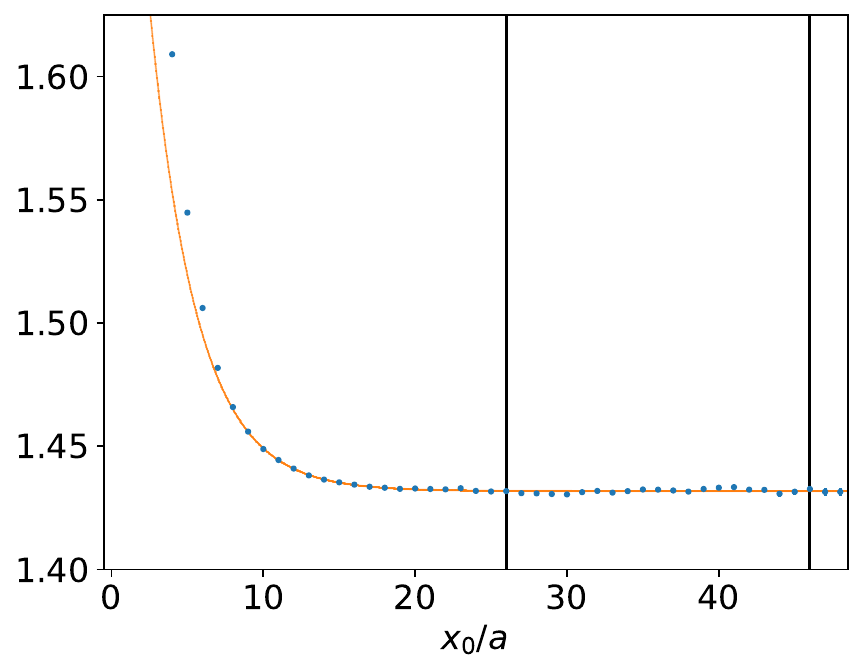}
    \end{tabular}
    \caption{Representative plateau in the mass of the $D^\ast_{\rm avg}$-meson (left) and the $J/\Psi$-meson (right), shown here for an ensemble with lattice spacing $0.094\,{\rm fm}$, in vicinity of the estimated charm quark mass parameter $\Delta_{\rm c}$.}
    \label{fig:vec_meson_plats}
\end{figure}

\begin{figure}
    \begin{tabular}{cc}
        \includegraphics[width = .45\linewidth]{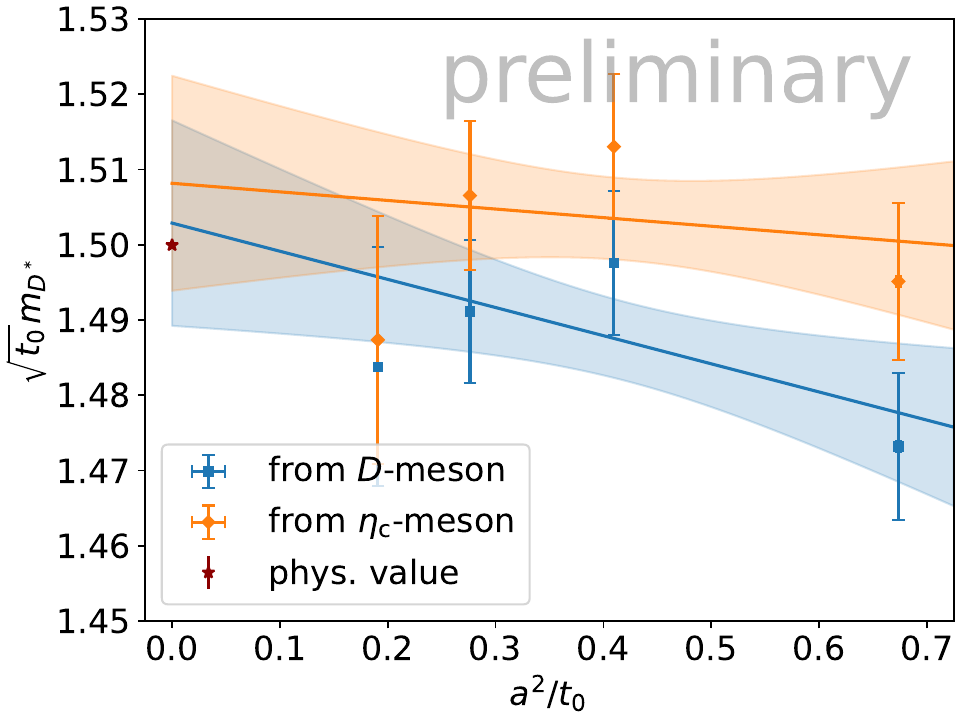}& 
         \includegraphics[width = .45\linewidth]{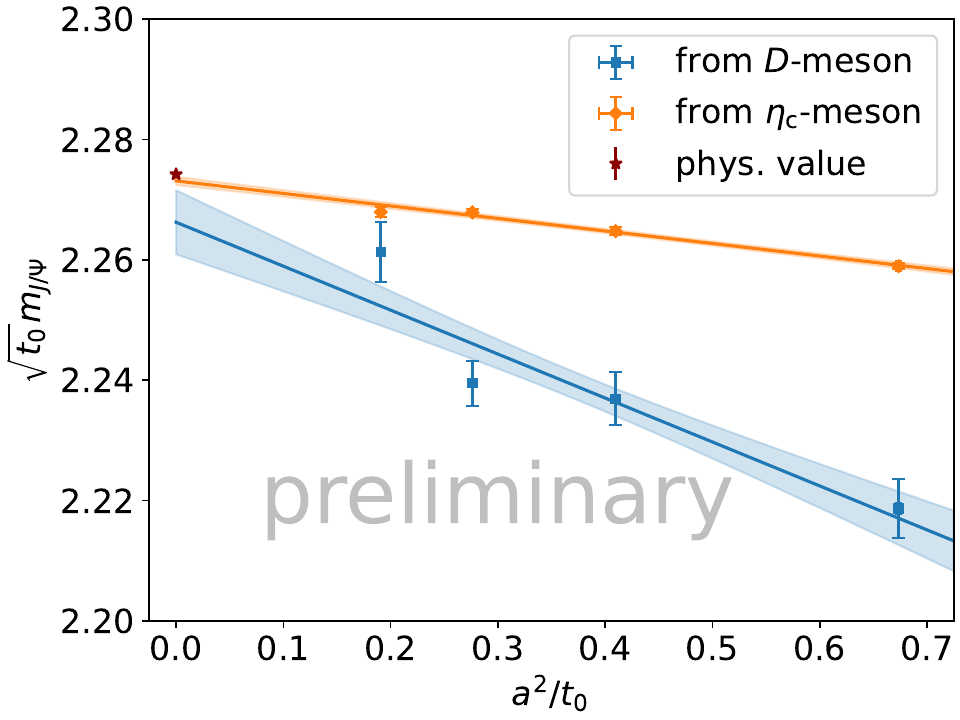}
    \end{tabular}
    \caption{Continuum limit extrapolations linear in $a^2/t_0$ of the $D^\ast_{\rm avg}$ (left) and $J/\Psi$ (right) mass, resulting from the two considered ways of fixing the charm quark mass scale. The physical value is also displayed for reference.}
    \label{fig:vec_meson_limits}
\end{figure}

After the charm quark mass has been fixed in both ways, we compare the outcomes. In fig.~\ref{fig:vec_meson_limits}, the continuum limits of the resulting masses in the vector channel are shown for the $J/\Psi$ charmonium state and the $D^\ast_{\rm avg}$-meson. In both plots, an overall error originating from the physical value of the scale parameter $t_0^{\rm phys}$ \cite{Bali2022} is still ignored, in order to test solely for the cut-off effects associated with the respective procedure to fix the charm quark mass parameter. Of course, in later applications of these fixtures, this error would have to be accounted for. The plateau averages entering here and exemplarily depicted in fig.~\ref{fig:vec_meson_plats} were obtained in the same manner as described above.
In particular, one observes again that the ground state effective mass of the charmonium state allows for a much wider plateau than the $D^\ast_{\rm avg}$-meson state.

\begin{figure}
\centering
    \includegraphics[width = .6\linewidth]{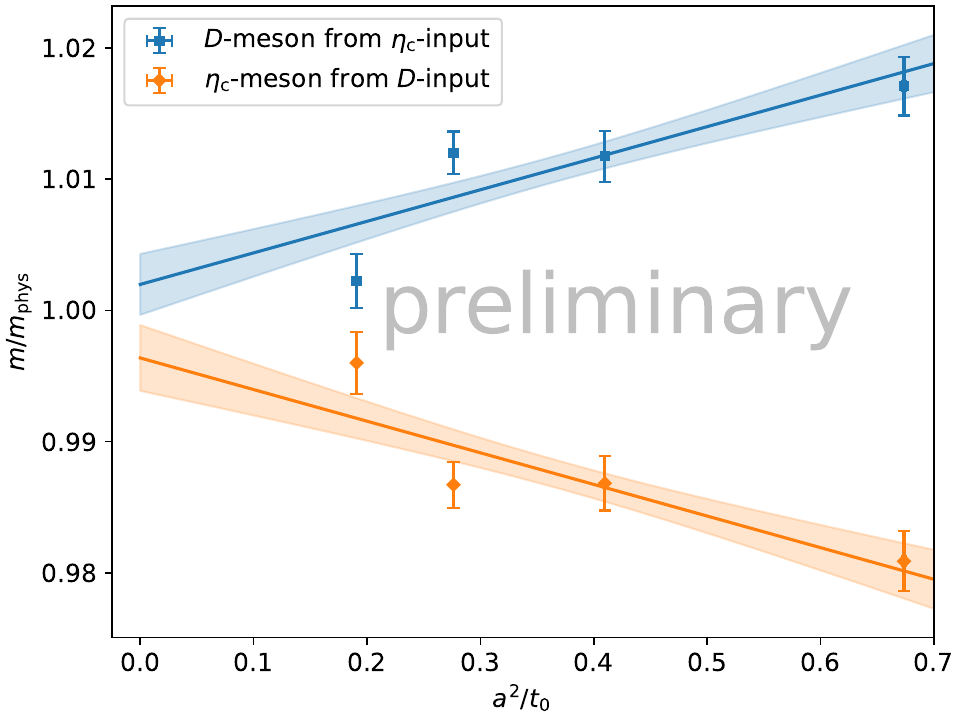}
    \caption{Relative cut-off effects emerging from the two considered ways of fixing the charm quark mass scale. Shown are mass ratios to the respective physical values, i.e., for the $D_{\rm avg}$-meson, where the charm quark's mass parameter was fixed through $\etac$-meson input (blue squares), and for the $\etac$-meson with the input being the $D_{\rm avg}$-meson mass (orange diamonds).}
    \label{fig:cont_limits_cross}
\end{figure}

In addition to the vector channel, we also address relative cut-off effects, which are illustrated in fig.~\ref{fig:cont_limits_cross}. There we display the outcome of a cross-evaluation of our methods: mass inputs and outputs are interchanged such that the mass of the $\etac$-meson is evaluated at the charm quark mass parameter fixed through the $D_{\rm avg}$-meson and vice versa. The graphs hint at slightly milder cut-off effects in the latter case.

\section{Outlook}
In this contribution, a first study in heavy quark physics in the framework of stabilised Wilson fermions was carried out and two ways to fix the charm quark mass for this setup have been investigated.

Next immediate steps are to perform multistate fits in order to further corroborate the determination of the ground-state plateau of the effective masses. This is not only particularly helpful on finer lattices and in other channels than the pseudo-scalar one, it would also directly enable the extraction of hadronic matrix elements (relevant for determining, e.g., leptonic decay constants), which are of interest to further explore phenomenology in the stabilised Wilson fermion framework.

Furthermore, we have seen that the continuum extrapolations of the mass channels under study are well controlled, especially in the case of the $J/\Psi$-meson, where the effective mass plateaux of the $\etac$- and the $J/\Psi$-meson are well-defined, show small cut-off effects and yield a continuum limit close to the expected physical value. Nevertheless, it is desirable to increase statistics, to add a smaller lattice spacing, and to apply small corrections to make the tuning of the simulated light-quark line of constant physics exact.
With these extensions we will probe the continuum limit once more, including charmed matrix elements, and check for potential higher-order cut-off effects.

In the long run, we are also interested in using the subtracted quark mass directly as a means of fixing the charm quark mass. However, as mentioned earlier, it is critical to have all renormalisation and improvement parameters under control, including the terms depending on ${\rm Tr}[M_{\rm q}]$ as they do not vanish in our current setup.

Since the interpolations in the (bare) valence quark mass parameter to the physical input target fixing the charm quark mass is very well controlled in our setup, it is possible to use the once implemented interpolation function instead of running an extra set of measurements at the estimated charm quark mass parameter. This facilitates using different determinations of the charm quark mass parameter, as there is no need for new calculations in case that one particular choice would be favoured.

\section*{Acknowledgements}
This work is supported by the Deutsche Forschungsgemeinschaft (DFG) through the Research Training Group “GRK 2149: Strong and Weak Interactions – from Hadrons to Dark Matter” (J.T.K. and J.H.). 
Calculations for this publication were performed on the HPC cluster PALMA II of the University of Münster, subsidised by the DFG (INST 211/667-1).
\bibliographystyle{JHEP}
\bibliography{myLibrary.bib}

\end{document}